%% file: main.tex
\documentclass[aps,prl,reprint,superscriptaddress,twocolumn,superscriptaddress,nofootinbib]{revtex4-1}
\usepackage{amsmath, amsthm, mathrsfs, amssymb, amsfonts, mathtools}
\usepackage{graphics}
\usepackage{hyperref}
\hypersetup{colorlinks=true, linkcolor=blue, citecolor=magenta, filecolor=magenta, urlcolor=cyan}

\usepackage{natbib} 
\input{macros}
\newcommand{\dd}{\mathrm{d}}

\DeclareMathAlphabet{\mathpzc}{OT1}{pzc}{m}{it}

\usepackage{orcidlink}

\begin{document}

\title{Nonlinear oscillations of strings and beams}

\author{Filip Ficek\hspace{0.233ex}\orcidlink{0000-0001-5885-7064}}
\email{filip.ficek@univie.ac.at}
\affiliation{University of Vienna, Faculty of Mathematics, Oskar-Morgenstern-Platz 1, 1090 Vienna, Austria}
\affiliation{University of Vienna, Gravitational Physics, Boltzmanngasse 5, 1090 Vienna, Austria}
\author{Maciej Maliborski\hspace{0.233ex}\orcidlink{0000-0002-8621-9761}}
\email{maciej.maliborski@univie.ac.at}
\affiliation{University of Vienna, Faculty of Mathematics, Oskar-Morgenstern-Platz 1, 1090 Vienna, Austria}
\affiliation{University of Vienna, Gravitational Physics, Boltzmanngasse 5, 1090 Vienna, Austria}
\affiliation{TU Wien, Institute of Analysis and Scientific Computing, Wiedner Hauptstraße 8–10, 1040 Vienna, Austria}
\date{\today}
\begin{abstract}
We investigate the time-periodic solutions to the nonlinear wave and beam equations and uncover their intricate, fractal-like structure. In particular, we identify a new class of large-energy solutions with complex mode compositions and propose a systematic framework for their analysis. A Floquet stability study reveals that this class contains solutions that are linearly stable, suggesting that they may play a significant role in the nonlinear dynamics of the systems.
\end{abstract}

\maketitle

\begin{figure}[t]
    \includegraphics[width=0.95\linewidth]{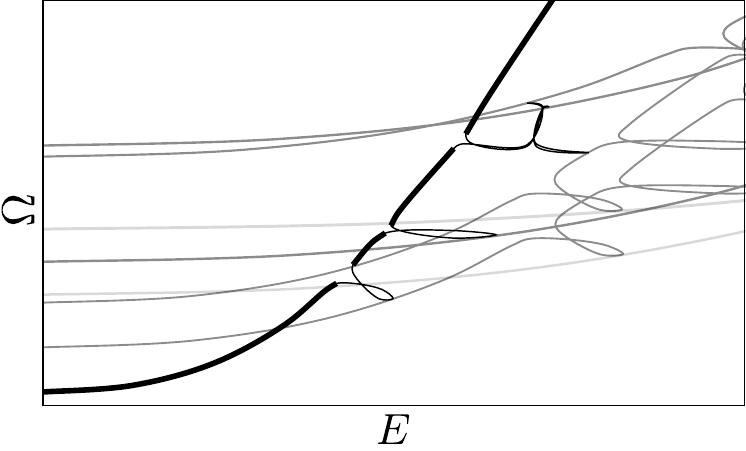}
    \caption{A simplified diagram presenting the structure of time-periodic solutions to Eq.~\eqref{eq:beam}. The thick black lines represent previously known family of solutions dominated by the $\cos\tau\, \sin x$ mode. Thinner black lines show structures formed by the new class of solutions described in this work. These structures connect to its own rescalings, depicted by gray lines, at bifurcation points. Analogously, these rescalings connect to further rescalings, illustrated with a lighter gray colour, leading to a fractal-like pattern. Note that the full picture would contain an infinite number of holes in the thick black curve filled with solutions belonging to the new class, and an infinite number of resacalings connected to each other.}
    \label{fig:cartoon}
\end{figure}

\textit{Introduction}---Wave and beam equations are among the simplest and most universal mechanical models. The former is central to fluid mechanics and the theory of elasticity, for example, describing vibrating strings.
Beyond continuum mechanics, the nonlinear wave equation plays a fundamental role in mathematical physics, particularly in field theory, where it models scalar field dynamics and nonlinear wave phenomena in contexts ranging from classical and quantum field theory to general relativity and cosmology.
The latter equation finds application in the modelling of vibrating rods \cite{bauchau2009euler,landau1986theory}, suspension bridges \cite{lazer1990large, HEUNGCHOI1999649}, or nanoelectromechanical systems \cite{schmid2016fundamentals, kacem2009nonlinear}. One of the most fundamental phenomena investigated in the context of these equations is nonlinear oscillations \cite{nayfeh2024nonlinear, berti2007nonlinear}.
In this work we discuss the intricate structure of time-periodic solutions to these equations equipped with defocusing cubic nonlinearity. Previous works focused on families of solutions bifurcating from the solutions to linearised equations \cite{berti2008cantor, lidskii1988periodic, GMP.2005, gentile2009periodic, Bambusi.2001, Arioli.2017, hall1970periodic, eliasson2014kam, LEE2000631}. These solutions are dominated by a single mode and form Cantor-like families, see thick lines in Fig.~\ref{fig:cartoon}. Here, we complement those results and show the existence of new classes of solutions represented by thinner lines in Fig.~\ref{fig:cartoon}. They are characterised by more complex mode compositions and their inclusion leads to an intricate, fractal-like pattern of the solutions. We also perform the first study of their stability. It turns out that some of them are linearly stable and could influence the long-time nonlinear evolution.
Even though we restrict ourselves to a defocusing cubic nonlinearity, similar results seem to hold for a much wider class of problems.

Both studied models can be described by a common equation
\begin{align}\label{eq:beam0}
    \partial_t^2u + (-1)^{\nu} \partial_x^{2\nu}u +u^3=0\,,
\end{align}
where $\nu=1$ gives the wave equation, while $\nu=2$ yields the beam equation.
We pose Eq.~\eqref{eq:beam0} on an interval $[0,\pi]$ and equip it with Dirichlet boundary conditions $u(t,0)=u(t,\pi)=0$ in the case of the wave equation and Navier boundary conditions
\begin{align*}
    u(t,0)=u(t,\pi)=\partial_x^2 u(t,0)=\partial_x^2 u(t,\pi)=0
\end{align*}
for the beam equation.

Time-periodic solutions are non-trivial solutions for which there is a period $T>0$ such that $u(t+T,x)=u(t,x)$. Since there is no external force, it is convenient to introduce the rescaled time $\tau=\Omega\,t$ so that the period becomes $2\pi$. Then $\Omega$ can be interpreted as the frequency of the solution $u$ and Eq.~\eqref{eq:beam0} becomes
\begin{align}\label{eq:beam}
    \Omega^2\partial_\tau^2u +(-1)^\nu\partial_x^{2\nu}u +u^3=0\,.
\end{align}
These models possess the conserved energies
\begin{align*}
    E[u]=\int_{0}^{\pi}\left(\frac{1}{2}\Omega^2\left(\partial_\tau u\right)^2+\frac{1}{2}\left(\partial_x^{\nu} u\right)^2+\frac{1}{4} u^4\right)\dd x\,.
\end{align*}

Note that Eq.~\eqref{eq:beam} admits a scaling symmetry: if $u(\tau,x)$ is a time-periodic solution with frequency $\Omega$, then
\begin{align}\label{eq:rescaling}
    u(\tau,x)\to n^{\nu}u(m\tau,nx)\,,\quad \Omega\to n^{\nu}\Omega/m
\end{align}
yields new solutions of Eq.~\eqref{eq:beam} for all positive integers $m,n$. The energy of the solution scales under this operation as $E\to n^{4\nu}E$. This property is particularly useful when comparing solutions emerging from different linear frequencies and analysing bifurcations.

\begin{figure*}[t!]
    \includegraphics[width=1.0\linewidth]{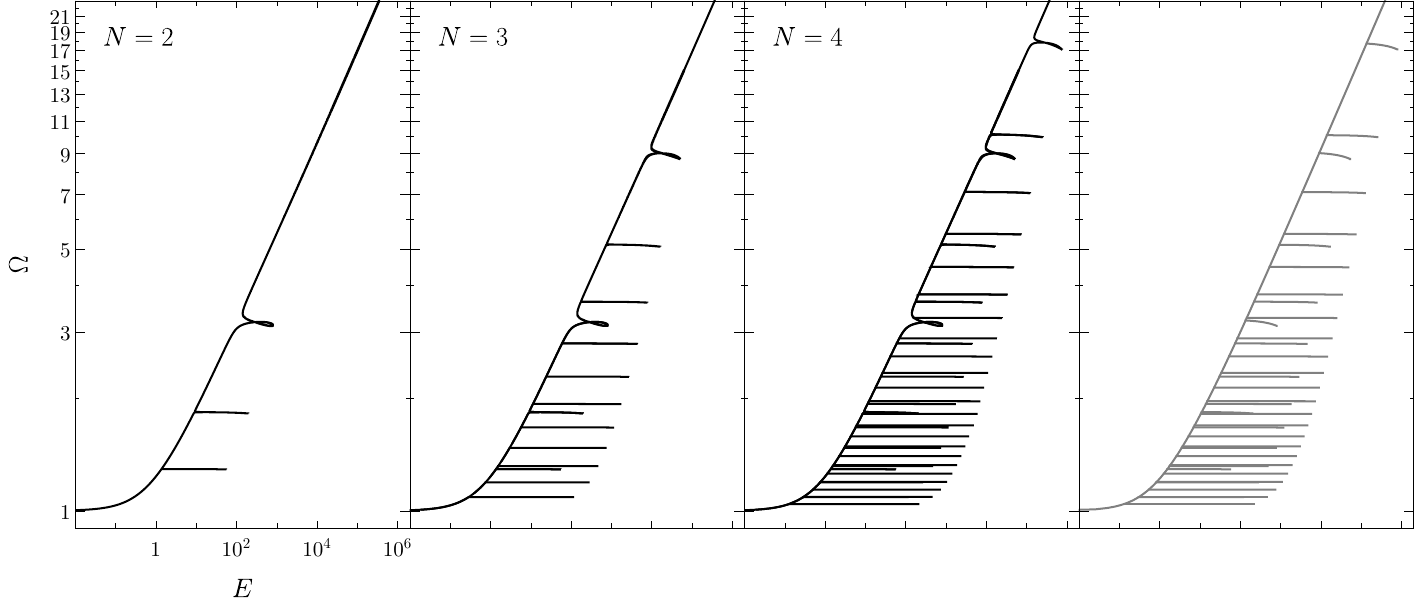}
    \caption{Energy–frequency $(E,\Omega)$ plots for solutions of the form \eqref{eq:ansatz} with increasing truncations $N$ for the beam equation ($\nu=2$). As $N$ increases, new solutions emerge as structures branching from the main curve. In the limit $N\to\infty$, this behaviour suggests an infinitely intricate solution structure for the corresponding PDE \eqref{eq:beam0}.
    For $N=4$, we also display the reducible tree, consisting of the trunk \eqref{eqn:reducible_solution_trunk} and all two-mode solutions \eqref{eqn:reducible_solution_2mode} (branches) available at this truncation (rightmost panel).
    }
    \label{fig:varN}
\end{figure*}

\textit{Galerkin scheme}---To find time-periodic solutions to Eq.~\eqref{eq:beam} we employ the Galerkin scheme. Let us fix some truncation numbers $M$, $N$ and write the solution $u$ as 
\begin{align}\label{eq:ansatz}
    u(\tau,x)=\sum_{m=0}^{M-1} \sum_{n=0}^{N-1} \hat{u}_{m,n} \cos(2m+1)\tau\, \sin(2n+1)x\,,
\end{align}
(thanks to the cubic nature of the nonlinearity, we can constrain ourselves to odd modes only). It is clear that such $u$ automatically satisfies the boundary conditions and is time-periodic with period $2\pi$. By plugging this ansatz into Eq.~\eqref{eq:beam} and projecting the resulting equation on $MN$ modes $\cos(2m+1)\tau\, \sin(2n+1)x$ present in the sum \eqref{eq:ansatz} we end up with the system of $MN$ algebraic equations. They involve $MN+1$ unknowns: $MN$ coefficients $\hat{u}_{m,n}$ and the frequency $\Omega$. We are looking for a continuous family of solutions bifurcating from zero at $\Omega=1$ and initially dominated by the fundamental mode $\cos\tau\,\sin x$ since families bifurcating from other frequencies $\Omega$ can be obtained from this one using \eqref{eq:rescaling}.
Since we expect this family to exhibit non-monotonic dependence on a parameter (due to folds and bifurcation points), instead of parametrising it with $\Omega$, we use the pseudo-arclength continuation method \cite{Keller.1987}. Using this parametrisation we solve numerically the succeeding algebraic equations and obtain mode compositions of the solutions from the considered family. For further details regarding the numerical scheme, we refer to \cite{Ficek.2024A}.

\begin{figure*}[t!]
    \includegraphics[width=1.0\linewidth]{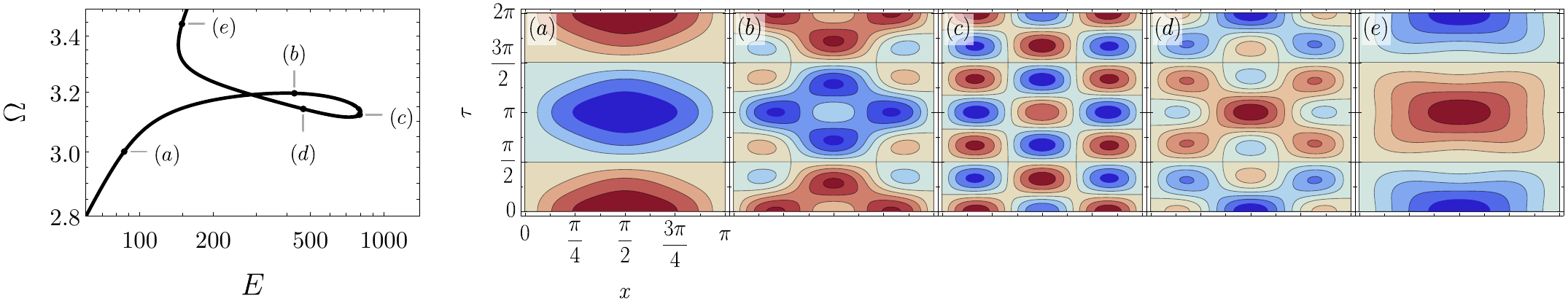}
    \caption{Density plots of selected solutions on one of the branches (the $N=2$ case). Blue and red colours show negative and positive values respectively.}
    \label{fig:profiles}
\end{figure*}

The results of the numerical calculations for the beam equation ($\nu=2$) under increasing truncations $N$ with $M=N^2$ are shown in Fig.~\ref{fig:varN}. (We note that further increasing $M$ does not qualitatively alter the solution structure; in particular, no new branches appear. This can be understood by studying reducible systems, see next section.)
We can see there the main curve bifurcating from zero along which the frequencies of the solutions increase steadily. All solutions described in the literature discussed in the introduction lie on this part, that we call the \textit{trunk}.
However, one can also observe solution \textit{branches} emerging from it. As $N$ increases, new branches appear with the already existing branches still present. In addition, one can also observe that some branches are prone to the emergence of subsequent, more complicated structures; see discussion below.

\begin{figure}[t!]
    \includegraphics[width=1.0\linewidth]{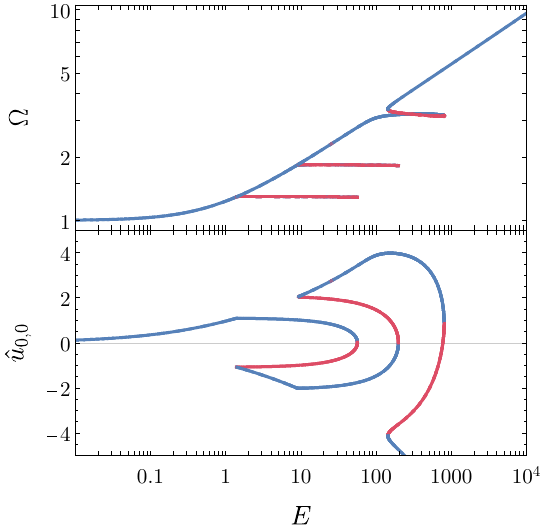}
    \caption{Linear stability of solutions for $N=2$ case for the beam equation (blue: stable, red: unstable). Upper panel: the $(E,\Omega)$ diagram. Lower panel: amplitude of the $\cos\tau\,\sin x$ mode, $\hat{u}_{0,0}$, versus $E$. The trunk is predominantly stable, with a small unstable segment near $(E,\Omega)\approx(27,2.3)$, while most noticeable stability changes occur on branches. The amplitude plot highlights that stability transition occur at the local extrema of energy. 
    }
    \label{fig:4}
\end{figure}

Next let us study profiles of the solutions lying on one of the branches and in its neighbourhood, as shown in Fig.~\ref{fig:profiles}. At the trunk, the solution is dominated by the fundamental mode $\cos\tau\,\sin x$. As we trace the solution curve as it enters the branch, we can see that the fundamental mode amplitude decreases while the amplitude of a certain higher mode increases ($\cos 3\tau\,\sin 3x$ in the case of the branch presented in Fig.~\ref{fig:profiles}). At the end of the branch, the fundamental mode is totally absent (see Fig.~\ref{fig:4}) and this new mode dominates. This is also a bifurcation point of a rescaled trunk dominated by this mode. As we go back to the trunk along the solution curve, we observe an opposite behaviour with the fundamental mode starting to dominate again.

Results for the wave equation ($\nu=1$) are qualitatively the same and can be found in \cite{Ficek.2024A}.

\textit{Reducible systems}---The observations just made may suggest trying to investigate the structure of the branches by considering interactions between just two modes, i.e., Galerkin schemes based on the ansatz
\begin{align}
    \label{eq:16.08.25_01}
    u(\tau,x)=A\cos\tau\,\sin x+B \cos (2m+1)\tau\,\sin (2n+1)x\,.
\end{align}
In the following, we assume that $m\geq1$, $n\geq 1$ and $(2m+1)<(2n+1)^{\nu}$, so the system of equations takes the form of
\begin{equation}
	\label{eqn:reducible_system}
	\left\{\begin{aligned}
		A\left[9A^2+12B^2-16\Omega^2+16\right]&=0\,,\\
		B\left[12A^2+9B^2-16(2m+1)^2\Omega^2+16(2n+1)^{2\nu}\right]&=0\,,
	\end{aligned}\right.
\end{equation}
with the exception of the case $(m,n)=(1,1)$ for $\nu=2$. Then the actual Galerkin system includes additional terms, see Supplemental Material, but \eqref{eqn:reducible_system} remains a satisfying approximation.
It can be solved explicitly, as it is linear in $A^2$ and $B^2$ (using the language introduced in \cite{Ficek.2024B}, such systems composed of minimally coupled modes are called \textit{reducible}). Among its solutions, we can find the single mode solution approximating the trunk,
\begin{align}
    \label{eqn:reducible_solution_trunk}
    A=\frac{4}{3}\sqrt{\Omega^2-1}\,,\qquad B=0\,,
\end{align}
and also a two-modes solution bifurcating from it
\begin{align}
    \label{eqn:reducible_solution_2mode}
    A&=\frac{4}{\sqrt{21}}\sqrt{\left[4(2m+1)^2-3\right]\Omega^2-\left[4(2n+1)^{2\nu}-3\right]}\,,\\
    B&=\frac{4}{\sqrt{21}}\sqrt{\left[3(2n+1)^{2\nu}-4\right]-\left[3(2m+1)^2-4\right]\Omega^2}\,.\nonumber
\end{align}
Such a solution is real only in the range of $\Omega$ satisfying
\begin{align*}
    \frac{4(2n+1)^{2\nu} -3}{4(2m+1)^2 -3}<\Omega^2< \frac{3(2n+1)^{2\nu} -4}{3(2m+1)^2 -4}\,.
\end{align*}
These two-modes solutions replicate the structure of the solutions, giving approximate locations, shapes, and dominant mode of all the branches up to the considered order of truncation for both wave ($\nu=1$), see \cite{Ficek.2024B}, and beam ($\nu=2$) equations, see Fig.~\ref{fig:varN}.

This approach gives us a systematic way of studying the structure of the branches. In particular, we can deduce that in the limit $N\rightarrow\infty$ the branches populate the trunks densely.
It agrees with the previous rigorous results predicting that the trunk is a Cantor-like set, at the same time suggesting that the resulting holes in frequencies of the solutions are filled by more energetic solutions.
It also implies the existence of solutions with arbitrarily large energies and frequencies arbitrarily close to $\Omega=1$.

\begin{figure}[t!]
    \centering
    \includegraphics[width=1.0\linewidth]{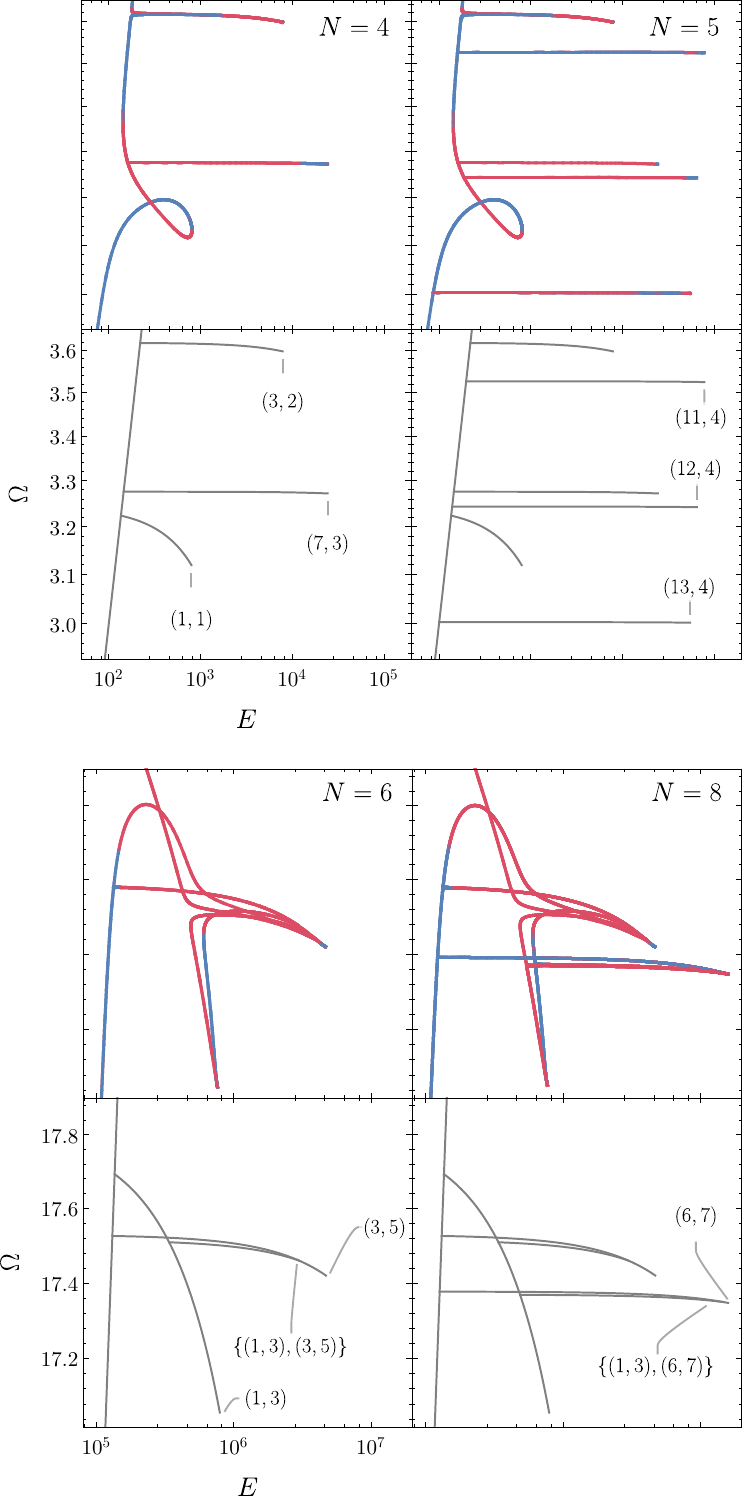}
    \caption{Energy–frequency $(E,\Omega)$ diagrams for various truncations $N$ with stability coding (blue: stable, red: unstable) and reducible trees (gray) for the beam equation ($\nu=2$).
    Mode numbers accompanying the fundamental mode are labelled as in \eqref{eq:16.08.25_01}.
    }
    \label{fig:partCpartD}
\end{figure}

A similar approach can be used to reproduce more complicated structures that bifurcate from the branches as the truncation $N$ increases. For example, by considering reducible systems spanned by appropriate sets of three modes, we get the linear system of equations similar to \eqref{eqn:reducible_system} that can also be explicitly solved. Then, combining their two-modes and three-modes solutions leads to patterns similar to the ones observed in the numerical solution of the full Galerkin system, see Fig.~\ref{fig:partCpartD}.
For more details, we refer to \cite{Ficek.2024B}.

Finally, recall the scaling symmetry of Eq.~\eqref{eq:beam}. As a consequence, the investigated family of solutions that bifurcate from $\Omega=1$ from the linearised solution dominated by the mode $\cos\tau\,\sin x$ leads to new families that are its appropriate rescalings. However, these families are not separated from the original solution curve. 
They connect with it at the ends of its branches (bifurcation points) via solutions with the absence of the fundamental mode.
This feature can be recreated by reducible systems \cite{Ficek.2024B}.

\textit{Linear stability}---We next investigate the linear stability of the constructed solutions using Floquet theory \cite{teschl2012ordinary}. By linearising Eq.~\eqref{eq:beam} around a periodic solution $u$ we get
\begin{align}\label{eq:pert}
    \Omega^2\partial_\tau^2 v +(-1)^\nu\partial_x^{2\nu}v +3u^2\,v=0\,.
\end{align}
We introduce a truncation to odd $K$ spatial modes and consider perturbations of the form $v(\tau,x)=\sum_{k=0}^{K-1} a_{k}(\tau)\sin (k+1)x$, whereas we approximate $u$ with the numerical solution in the form \eqref{eq:ansatz} with truncation $N=(K+1)/2$ and $M$ large enough for the residue to be small.
Projecting Eq.~\eqref{eq:pert} into $K$ spatial modes we get a Hamiltonian system of size $2K$
\begin{align*}
\frac{\dd}{\dd \tau}\begin{pmatrix} p\\q \end{pmatrix}=
\begin{pmatrix} 0 & I \\ -L/\Omega^2 & 0\end{pmatrix} \begin{pmatrix} p\\q \end{pmatrix}\,,
\end{align*}
where $p=(a_0,...,a_{K-1})^T$ and $q=(\dot{a}_0,...,\dot{a}_{K-1})^T$, and $L$ is a linear operator encoding how $(-1)^{\nu+1}\partial_x^{2\nu}-3u^2$ acts on the modes $\sin(k+1)x$. We evolve this system numerically using a high-order symplectic integrator \cite{Hairer.2006_book} for $2K$ independent unit-vector initial conditions over the period of $2\pi$. Then the resulting vectors form the columns of the $2K \times 2K$ monodromy matrix, whose eigenvalues $\lambda$ are the Floquet multipliers encoding the linear stability information for $u$. Since the system is Hamiltonian, it admits neither attractors nor repellers; hence, if an eigenvalue satisfies $|\lambda| < 1$, there must be a counterpart with $|\lambda| > 1$, implying linear instability; all $\lambda$ on the unit circle indicates linear stability.

Figure~\ref{fig:4} displays the $(E,\Omega)$ diagram for $N=2$ (cf. Fig.~\ref{fig:varN}), as well as $(E,\hat{u}_{0,0})$ for clarity, colour-coded by stability. Stable solutions---identified via $||\lambda|-1|\leq 10^{-12}$---occur not only along the trunk but also on branches.
Figure \ref{fig:partCpartD} zooms into two regions of the $(E,\Omega)$ diagram for higher truncations, showing in greater detail qualitative agreement with reducible-system predictions and revealing stable regions on new higher-order structures.

Overall, these results indicate that the newly discovered class of solutions contains linearly stable members, which may play an important role in the nonlinear dynamics.
The details of the employed method, together with an alternative approach and the results for the wave equation, can be found in the upcoming work \cite{Ficek.2025D}.

\textit{Summary}---We have uncovered an intricate, fractal-like structure of time-periodic solutions to the defocusing cubic wave and beam equations on a bounded interval.
In addition to the well-known Cantor-type solution families established rigorously at small amplitude, we have found a new class of large-energy, multi-mode solutions that populate the gaps of these Cantor sets and connect their rescaled copies.
We propose a systematic framework for their analysis, based on what we call reducible systems, which accurately models the complex web of solutions and their rescalings \cite{Ficek.2024B}.
We have shown that some members of this class are linearly stable, indicating their potential relevance in the nonlinear dynamics. 
Recently, in \cite{Ficek.2025B} we provided a computer-assisted proof of existence for solutions of this type for the wave equation, inspired by \cite{Arioli.2017}, and an extension of the proof to the beam equation appears straightforward.
Our results bridge previous rigorous existence results with the finite-amplitude regime and suggest that similar structures may arise in nonlinear field theories on bounded domains, including self-gravitating or self-interacting scalar fields in a cavity \cite{Maliborski.2012} and with negative cosmological constant \cite{Maliborski.2013, Ficek.2025}, as well as in physical systems modelled by the equations \eqref{eq:beam0}---such as finite strings, suspension bridges, and nanoelectromechanical resonators---where they may be experimentally observable and influence long-time behavior.

\vspace{1ex}

\textit{Acknowledgments}---We acknowledge the support of the Austrian Science Fund (FWF) through Project \href{http://doi.org/10.55776/P36455}{P 36455}, the START-Project \href{http://doi.org/10.55776/Y963}{Y 963}, and the Wittgenstein Award \href{http://doi.org/10.55776/Z387}{Z 387}.

\bibliography{lib} 

\onecolumngrid

\clearpage
\begin{center}
\textbf{\large Supplemental Material}
\end{center}

\setcounter{section}{0}
\renewcommand{\thesection}{S\arabic{section}}

\section{Nonreducible Galerkin system}
\begin{figure}[h]
    \centering
    \includegraphics[width=0.85\textwidth]{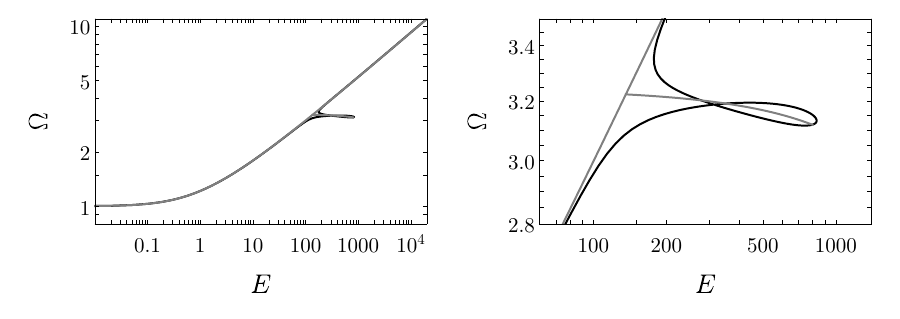}
    \caption{Comparison of the solutions to Eq.~\eqref{eqn:reducible_system} with $(m,n)=(1,1)$ and $\nu=2$ (gray) and numerical solution to Eq.~\eqref{eqn:nonreducible_system} (black).}
    \label{fig:nonreducible}
\end{figure}

In the main text, we have used the reducible system \eqref{eqn:reducible_system} to describe the solutions to the Galerkin system spanned by the modes $\cos\tau\,\sin x$ and $\cos 3\tau\,\sin 3x$ for the beam equation. However, in this case, the actual Galerkin system has the form
\begin{align}
	\label{eqn:nonreducible_system}
	\left\{\begin{aligned}
		A\left[9A^2+12B^2-16\Omega^2+16\right] -3A^{2}B&=0\,,\\
		B\left[12A^2+9B^2-144\Omega^2+1296\right]-A^3&=0\,,
	\end{aligned}\right.
\end{align}
so \eqref{eqn:reducible_system} is only an approximation obtained by dropping the last terms.
We compare solutions to \eqref{eqn:reducible_system} and \eqref{eqn:nonreducible_system} in Fig.~\ref{fig:nonreducible}. 

Let us point out that a non-reducible two-modes system leading to a branch is a feature introduced by the beam equation. In \cite{Ficek.2024B} we have proven that for the wave equation only reducible systems give raise to branches emerging from the trunk. One can use similar methods to show that for the beam equation the only non-reducible two-modes system producing a branch is the one discussed here.

\end{document}

%% file: macros.tex
\usepackage[dvipsnames]{xcolor}
\usepackage{soul}
\usepackage[colorinlistoftodos, textwidth=1.75cm]{todonotes}
\usepackage[color, final]{showkeys}
\reversemarginpar
\usepackage[normalem]{ulem} 

\usepackage{lmodern}

\def\margincomentfontsize{\tiny}
\newcounter{todocounter}
\newcommand{\todonum}[2][]{
        \stepcounter{todocounter}\todo[#1]{\thetodocounter: #2}
}

\definecolor{refkey}{gray}{.75}
\definecolor{labelkey}{gray}{.55}

\def\1{\mathbb{1}} 

\makeatletter
\if@todonotes@disabled
\newcommand{\error}[1]{}
\newcommand{\warning}[1]{}
\newcommand{\comment}[1]{}

\else

\newcommand{\error}[1]{\todonum[color=RedOrange,size=\margincomentfontsize]{\ttfamily{#1}}}
\newcommand{\warning}[1]{\todonum[tickmarkheight=0.5ex, color=TealBlue,size=\margincomentfontsize]{\ttfamily{#1}}}
\newcommand{\comment}[1]{\todonum[color=Goldenrod,size=\margincomentfontsize]{\ttfamily{#1}}}

\fi
\makeatother

\newcommand{\editremove}[1]{}
\newcommand{\edittrash}[1]{}

\colorlet{lightblue}{Cyan!40!White}

\colorlet{lightpink}{Magenta!50!White}

\colorlet{lightorange}{Orange!60!White}




%% file: main.bbl
\begin{thebibliography}{27}
\providecommand{\natexlab}[1]{#1}
\providecommand{\url}[1]{\texttt{#1}}
\expandafter\ifx\csname urlstyle\endcsname\relax
  \providecommand{\doi}[1]{doi: #1}\else
  \providecommand{\doi}{doi: \begingroup \urlstyle{rm}\Url}\fi

\bibitem[Bauchau and Craig(2009)]{bauchau2009euler}
Oliver~A. Bauchau and James~I. Craig.
\newblock {Euler-Bernoulli beam theory}.
\newblock In Oliver~A Bauchau and James~I Craig, editors, \emph{{Structural Analysis}}, Solid Mechanics and Its Applications, pages 173--221. Springer Dordrecht, 2009.
\newblock \doi{10.1007/978-90-481-2516-6_5}.

\bibitem[Landau and Lifshitz(1986)]{landau1986theory}
Lev~D. Landau and Evgeny~M. Lifshitz.
\newblock \emph{Theory of Elasticity}, volume~7 of \emph{Course of Theoretical Physics}.
\newblock Butterworth-Heinemann, 3rd edition, 1986.
\newblock ISBN 978-0-7506-2633-0.

\bibitem[Lazer and McKenna(1990)]{lazer1990large}
Alan~C. Lazer and Patrick~J. McKenna.
\newblock {Large-Amplitude Periodic Oscillations in Suspension Bridges: Some New Connections with Nonlinear Analysis}.
\newblock \emph{Siam Review}, 32\penalty0 (4):\penalty0 537--578, 1990.
\newblock \doi{10.1137/1032120}.

\bibitem[Choi and Jung(1999)]{HEUNGCHOI1999649}
Q.~Heung Choi and Tacksun Jung.
\newblock {A nonlinear suspension bridge equation with nonconstant load}.
\newblock \emph{Nonlinear Analysis: Theory, Methods \& Applications}, 35\penalty0 (6):\penalty0 649--668, 1999.
\newblock ISSN 0362-546X.
\newblock \doi{10.1016/S0362-546X(97)00616-0}.

\bibitem[Schmid et~al.(2016)Schmid, Villanueva, and Roukes]{schmid2016fundamentals}
Silvan Schmid, Luis~Guillermo Villanueva, and Michael~Lee Roukes.
\newblock \emph{Fundamentals of nanomechanical resonators}, volume~49.
\newblock Springer, 2016.

\bibitem[Kacem et~al.(2009)Kacem, Hentz, Pinto, Reig, and Nguyen]{kacem2009nonlinear}
Najib Kacem, Sebastien Hentz, David Pinto, Bruno Reig, and V~Nguyen.
\newblock {Nonlinear dynamics of nanomechanical beam resonators: improving the performance of NEMS-based sensors}.
\newblock \emph{Nanotechnology}, 20\penalty0 (27):\penalty0 275501, 2009.
\newblock \doi{10.1088/0957-4484/20/27/275501}.

\bibitem[Nayfeh and Mook(2024)]{nayfeh2024nonlinear}
Ali~H Nayfeh and Dean~T Mook.
\newblock \emph{{Nonlinear Oscillations}}.
\newblock John Wiley \& Sons, 2024.
\newblock ISBN 3527617590.

\bibitem[Berti(2007)]{berti2007nonlinear}
Massimiliano Berti.
\newblock \emph{{Nonlinear oscillations of Hamiltonian PDEs}}, volume~74 of \emph{Progress in Nonlinear Differential Equations and Their Applications}.
\newblock Birkhäuser, Boston, 2007.
\newblock \doi{10.1007/978-0-8176-4681-3}.

\bibitem[Berti and Bolle(2008)]{berti2008cantor}
Massimiliano Berti and Philippe Bolle.
\newblock {Cantor families of periodic solutions for wave equations via a variational principle}.
\newblock \emph{Advances in Mathematics}, 217\penalty0 (4):\penalty0 1671--1727, 2008.
\newblock \doi{https://doi.org/10.1016/j.aim.2007.11.004}.

\bibitem[Lidskii and Shul'man(1988)]{lidskii1988periodic}
Boris~Viktorovich Lidskii and Evgenii~Iosifovich Shul'man.
\newblock {Periodic solutions of the equation $u_{tt} - u_{xx} + u^3 = 0$}.
\newblock \emph{Functional Analysis and Its Applications}, 22\penalty0 (4):\penalty0 332--333, 1988.
\newblock ISSN 0016-2663.
\newblock \doi{10.1007/bf01077432}.

\bibitem[Gentile et~al.(2005)Gentile, Mastropietro, and Procesi]{GMP.2005}
Guido Gentile, Vieri Mastropietro, and Michela Procesi.
\newblock {Periodic Solutions for Completely Resonant Nonlinear Wave Equations with Dirichlet Boundary Conditions}.
\newblock \emph{Communications in Mathematical Physics}, 256:\penalty0 437--490, 2005.
\newblock ISSN 0010-3616.
\newblock \doi{10.1007/s00220-004-1255-8}.

\bibitem[Gentile and Procesi(2009)]{gentile2009periodic}
Guido Gentile and Michela Procesi.
\newblock Periodic solutions for a class of nonlinear partial differential equations in higher dimension.
\newblock \emph{Communications in Mathematical Physics}, 289\penalty0 (3):\penalty0 863--906, 2009.

\bibitem[Bambusi and Paleari(2001)]{Bambusi.2001}
Dario Bambusi and Simone Paleari.
\newblock {Families of Periodic Solutions of Resonant PDEs}.
\newblock \emph{Journal of Nonlinear Science}, 11\penalty0 (1):\penalty0 69--87, 2001.
\newblock ISSN 0938-8974.
\newblock \doi{10.1007/s003320010010}.

\bibitem[Arioli and Koch(2017)]{Arioli.2017}
Gianni Arioli and Hans Koch.
\newblock {Families of Periodic Solutions for Some Hamiltonian PDEs}.
\newblock \emph{SIAM Journal on Applied Dynamical Systems}, 16\penalty0 (1):\penalty0 1--15, 2017.
\newblock \doi{10.1137/16m1070177}.

\bibitem[Hall(1970)]{hall1970periodic}
William~S Hall.
\newblock Periodic solutions of a class of weakly nonlinear evolution equations.
\newblock \emph{Archive for Rational Mechanics and Analysis}, 39\penalty0 (4):\penalty0 294--322, 1970.

\bibitem[Eliasson et~al.(2014)Eliasson, Grebert, and Kuksin]{eliasson2014kam}
Hakan~L Eliasson, Benoit Grebert, and Sergei~B Kuksin.
\newblock Kam for the nonlinear beam equation 1: small-amplitude solutions.
\newblock \emph{arXiv preprint arXiv:1412.2803}, 2014.

\bibitem[Lee(2000)]{LEE2000631}
Cheng Lee.
\newblock Periodic solutions of beam equations with symmetry.
\newblock \emph{Nonlinear Analysis: Theory, Methods \& Applications}, 42\penalty0 (4):\penalty0 631--650, 2000.
\newblock ISSN 0362-546X.
\newblock \doi{10.1016/S0362-546X(99)00119-4}.

\bibitem[Keller(1987)]{Keller.1987}
Herbert~B. Keller.
\newblock \emph{{Lectures on Numerical Methods in Bifurcation Problems}}.
\newblock Tata Institute Lectures on Mathematics and Physics. Springer-Verlag, Berlin, 1987.
\newblock ISBN 3540183671.

\bibitem[Ficek and Maliborski(2025{\natexlab{a}})]{Ficek.2024A}
Filip Ficek and Maciej Maliborski.
\newblock {Periodic solutions for the 1D cubic wave equation with Dirichlet boundary conditions}.
\newblock \emph{Nonlinearity}, 38\penalty0 (6):\penalty0 065016, 2025{\natexlab{a}}.
\newblock \doi{10.1088/1361-6544/add831}.

\bibitem[Ficek and Maliborski(2024)]{Ficek.2024B}
Filip Ficek and Maciej Maliborski.
\newblock {Trees, trunks, and branches - bifurcation structure of time-periodic solutions to $u_{tt}-u_{xx}\pm u^3=0$}.
\newblock \emph{arXiv}, 2024.
\newblock \doi{10.48550/arxiv.2408.05158}.

\bibitem[Teschl(2012)]{teschl2012ordinary}
Gerald Teschl.
\newblock \emph{{Ordinary Differential Equations and Dynamical Systems}}, volume 140 of \emph{{Graduate Studies in Mathematics}}.
\newblock American Mathematical Soc., 2012.
\newblock \doi{10.1090/gsm/140}.

\bibitem[Hairer et~al.(2006)Hairer, Lubich, and Wanner]{Hairer.2006_book}
Ernst Hairer, Christian Lubich, and Gerhard Wanner.
\newblock \emph{Geometric Numerical Integration: Structure-Preserving Algorithms for Ordinary Differential Equations}.
\newblock Springer Series in Computational Mathematics. Springer Berlin, Heidelberg, 2006.
\newblock ISBN 978-3-540-30663-4.
\newblock \doi{10.1007/3-540-30666-8}.

\bibitem[Ficek and Maliborski(2025{\natexlab{b}})]{Ficek.2025D}
Filip Ficek and Maciej Maliborski.
\newblock {In preparation}, 2025{\natexlab{b}}.

\bibitem[Ficek and Maliborski(2025{\natexlab{c}})]{Ficek.2025B}
Filip Ficek and Maciej Maliborski.
\newblock {New class of time-periodic solutions to the 1D cubic wave equation}.
\newblock \emph{arXiv}, 2025{\natexlab{c}}.
\newblock \doi{10.48550/arXiv.2506.10839}.

\bibitem[Maliborski(2012)]{Maliborski.2012}
Maciej Maliborski.
\newblock {Instability of Flat Space Enclosed in a Cavity}.
\newblock \emph{Physical Review Letters}, 109\penalty0 (22):\penalty0 221101, 2012.
\newblock ISSN 0031-9007.
\newblock \doi{10.1103/physrevlett.109.221101}.

\bibitem[Maliborski and Rostworowski(2013)]{Maliborski.2013}
Maciej Maliborski and Andrzej Rostworowski.
\newblock {Time-Periodic Solutions in an Einstein AdS–Massless-Scalar-Field System}.
\newblock \emph{Physical Review Letters}, 111\penalty0 (5):\penalty0 051102, 2013.
\newblock ISSN 0031-9007.
\newblock \doi{10.1103/physrevlett.111.051102}.

\bibitem[Ficek and Maliborski(2025{\natexlab{d}})]{Ficek.2025}
Filip Ficek and Maciej Maliborski.
\newblock {Complex structure of time-periodic solutions decoded in Poincar\'e-Lindstedt series: the cubic conformal wave equation on $\mathbb{S}^3$}.
\newblock \emph{Physica D: Nonlinear Phenomena}, page 134864, 2025{\natexlab{d}}.
\newblock ISSN 0167-2789.
\newblock \doi{10.1016/j.physd.2025.134864}.

\end{thebibliography}
